\documentclass[a4paper,11pt]{article}
\usepackage{pos}

\newcommand{\ie}{\textit{i.e.}} 
\def\Journal#1#2#3#4{{#1}\,{#2}, #3 (#4);}
\newcommand{\etal}{et al.}
\def\citep#1{\cite{#1}}
\newcommand{\tableline}{\hline}
\newcommand{\DT}{\ensuremath{\Delta{T}}}
\newcommand{\R}{\ensuremath{R}}
\newcommand{\p}{{p}}
\newcommand{\He}{{He}}
\newcommand{\C}{{C}}
\newcommand{\Oxy}{{O}}

\newcommand{\ApJ}{Astrophys. J.}
\newcommand{\AeA}{Astron. \& Astrophys.}
\newcommand{\PRL}{Phys. Rev. Lett.}
\newcommand{\PRD}{Phys. Rev. D}
\newcommand{\PRC}{Phys. Rev. C}
\newcommand{\ASR}{Adv. Space Res.}

\newcommand{\JGR}{J. Geophys. Res.}

\renewcommand{\citet}{\cite}

\title{New insights from cross-correlation studies between Solar activity and Cosmic-ray fluxes}
\ShortTitle{New insights from cross-correlation studies between Solar activity and Cosmic-ray fluxes}

\author{Nicola Tomassetti}
\author{Bruna Bertucci}
\author{Emanuele Fiandrini}
\affiliation{Department of Physics and Earth's Science, Universit\`a degli Studi di Perugia, Italy}
\emailAdd{nicola.tomassetti@unipg.it}

\abstract{
  The observed variability of the cosmic-ray intensity in the interplanetary space is driven by the evolution of the Sun's magnetic activity over its 11-year quasiperiodical cycle. Investigating the relationship between solar activity indices and cosmic-ray intensity measurements is then essential for understanding the fundamental processes of particle transport in the heliosphere. Here we have performed a global characterization the solar modulation of cosmic rays over the solar activity cycle and for different energies of the cosmic particles. We present our cross-correlation studies using data from space experiments, neutron monitors and solar observatories collected over several solar cycles.
}
\FullConference{37$^{\rm{th}}$ International Cosmic Ray Conference (ICRC 2021)\\
		July 12th -- 23rd, 2021\\
		Online -- Berlin, Germany}
\begin{document}

\maketitle

\section{Introduction}      
\label{Sec::Introduction}   

Understanding the time-dependent relationship between the Sun's variability and Galactic cosmic-ray (GCR) flux modulation is essential
for the investigation of the GCR transport processes in the heliosphere, as well as for the establishment of predictive models of
GCR radiation in the interplanetary space. 
Predicting the level GCR radiation near-Earth is essential for astronauts and the electronic
components radiation hazard in long-duration space missions.
Also, understanding the GCR modulation and its underlying physical mechanisms
is crucial to investigate the physics of charged particle transport in magnetized plasmas, to  identify the sources of GCRs and, eventually,
to unveil the origin of GCR antimatter and cosmological dark matter.

The propagation of GCRs in the heliosphere is caused by the interactions of these particles with
magnetic turbulence and Solar wind disturbances which, in turn, are driven by the quasiperiodical cycle of solar activity.
A known proxy for the Sun's magnetic activity is the monthly-averaged number of dark spots appearing on its surface, the so-called sunspot number (SSN). 
The known anti-correlation between GCR flux and SSN has been investigated for long time in a large number of studies.
A remarkable feature of such a correlation is that the GCR flux variations appears to be delayed by several months with respect to the SSN variation.
The existence of a time lag between GCRs and solar activity may reflect a causal relationship among the two phenomena,
but its origin is still subjected to investigation.
In this work, we have reconstructed the temporal evolution of the modulation lag over five solar cycles. 
We used of a large collection of measurements from space-borne GCR detectors, neutron monitors, and solar observatories.
As we discuss, the investigation presented here may give important insights of the physical mechanisms of GCR transport in the heliosphere.

\section{The Solar Modulation of Galactic Cosmic Rays}  
\label{Sec::SolarModulation}                            

Inside the heliosphere, Galactic cosmic rays (GCRs) travel through a turbulent flowing plasma, \ie, the solar wind in its embedded magnetic field.
GCRs are subjected to several physical processes such as spatial diffusion on the small-scale irregularities of the turbulent magnetic field,
drift motion caused by gradient and curvature of the large-scale interplanetary magnetic field, advection and adiabatic deceleration over the outwardly expanding
solar wind plasma, and even diffusive reacceleration in proximity of the termination shock.
All these processes cause a significant variation of the energy spectra of GCRs. In particular, the spectrum of GCRs observed near-Earth
is significantly different from the one in interstellar space (IS) beyond the boundaries of the heliosphere.
Moreover, the GCR flux inside the heliosphere is subjected to a remarkable time-dependence 
which is known as \emph{solar modulation} effect \cite{Potgieter2013}.
Interestingly, the temporal evolution of the solar modulation effect follows a quasi-periodical behavior
which appears to be well correlated with the 11-year cycle of solar activity.
The solar cycle is related with generation strong magnetic fields in the interior of the
Sun and manifested by the periodical appearance of sunspots.
The observed SSN, also known as Wolf number and often reported on monthly basis,
is then a good proxy for the magnetic activity in the Sun, and it is widely used to characterize the phases of the 11-year solar cycle.
In this work, the SSN is used for investigating the anti-correlation relation between GCR flux and solar activity \cite{Usoskin1998,Ross2019}.

In this respect, a big challenge is to establish reliable relations between the GCR modulation effect and the Sun's variability.
Several physical models have been developed in order to predict the evolution of the GCR intensity in the inner heliosphere \citep{Norbury2018,Slaba2020,Kuznetsov2017,Matthia2013,Fiandrini2021}.
Investigating  solar modulation and its underlying physical mechanisms is central in astrophysics.
Moreover, the varying level of the GCR radiation provides a significant challenge for
space missions and air travelers. Thus, predicting the level GCR radiation near-Earth is essential for
astronauts and the electronic components radiation hazard in long-duration space missions.

To understand the  dynamics of GCR modulation and its connection with the manifestations of solar activity,
an important aspect is the observation of a \emph{time lag} between the two phenomena.
Observationally, several studies reported time lags of few months between the long-term variations of monthly SSN and NM counting rates.
The effect is seen in different indices characterizing the conditions in the heliosphere such as tilt angle or polar magnetic field.
For Solar Cycle 23, a lag of 8.1$\pm$0.1 months was established using GCR spectra data from space \citep{Tomassetti2017TimeLag}.
Theoretically, the lag can be interpreted in terms of the dynamics of the modulation region, \ie, the heliosphere. 
According to this picture, GCR observed near-Earth reflect the turbulence condition of the heliospheric plasma,
which is ejected by the Sun and expands radially.
Thus, for a wind of finite speed $V\sim$450\,km/s, and a modulation region of size $r_{0}\sim$\,120 AU,
it would takes $\tau\sim{V}/r_{0}\sim$\,one year to fill the region with the magnetic turbulence \citet{DormanDorman1967}. 
There may be, however, other explanations involving delayed response of GCRs to changes in the background plasma.
Furthermore there is no experimental consensus on the lag value, as it ranges from 0 to 18 months, depending on epochs and cycles.
Such a discrepancy, along with a reported odd-even dependence on the cycle number, 
suggests that the time lag has a regular quasi-periodical evolution with time \citep{Usoskin2001}.

\section{The data we used}  
\label{Sec::Data}           

To investigate the evolution of the modulation lag over the solar cycle,
we made use of several time-series of multichannel measurements from space missions, ground-based detectors, or solar observatories.
Solar indices such as the SSN are constantly monitored by
the SILSO/SIDC data center of the \emph{Royal Observatory of Belgium} \cite{CletteLefevre2016}. 
From these data, we built a smoothed $\hat{S}(t)$-function that interpolates the time-series of monthly SSN.
For the temporal evolution of the GCR flux we used direct and indirect measurements.
Direct measurements of particle- and energy-resolved GCR fluxes have been done by long-running experiments such as
the MED on IMP-8 spacecraft (since 1972 to 2000 \citet{McDonald2003}),
the HET telescopes on Voyager-1 and -2 space probes (1979-present, \citet{Cummings2016})
CRIS on ACE (1997-present \citet{Wiedenbeck2009}),
EPHIN on SOHO (1995-2005 \citet{Kuhl2016}),
and more recently the magnetic spectrometers
PAMELA (on satellite Resurs-DK1, 2006-2014 \citet{Adriani2013,Martucci2018})
and AMS-02 (on the International Space Station, 2011-present \citet{Aguilar2018PHeVSTime,Aguilar2015Proton,Aguilar2015Helium}.
In this work however we are mostly focused on indirect measurements.
In particular,  we use time-series of monthly-averaged rates from neutron monitor (NM) detectors,
which continuously measure the time variation of GCRs since several decades \cite{Vaisanen2021,Mavromichalaki2011}.
We have considered the stations in Oulu, Kiel, Newark, Moscow, Jungfraujoch, and Rome.
For these stations we consider data from 1965 and 2020, covering Solar Cycle No. 20, 21, 22, 23, and 24.
The basic properties of the considered stations are listed in Table\,\ref{Tab::NMStations}. 
\setlength{\tabcolsep}{0.036in} 
\begin{table*}[!t]
\begin{center}
\small
\begin{tabular}{cccccc}
\tableline
\tableline
NM station & \href{http://www01.nmdb.eu/station/newk/}{NEWK} & \href{http://www01.nmdb.eu/station/oulu/}{OULU} &  \href{http://www01.nmdb.eu/station/kiel/}{KIEL} & \href{http://www01.nmdb.eu/station/jung}{JUNG} & \href{http://www01.nmdb.eu/station/rome}{ROME}  \\
\tableline
Detector type     & 9-NM64              & 9-NM64               & 18-NM64              & 3-NM64            & 20-NM64\\
Location          & Newark US           & Oulu FI              & Kiel DE              & Jungfraujoch CH   & Rome IT\\
Coordinates       & 39.68\,N 75.75\,W   & 65.05\,N, 25.47\,E   & 54.34\,N, 10.12\,E   & 46.55\,N, 7.98\,E & 41.86\,N, 12.47\,E \\
Altitude          & 50\,m               & 15\,m                & 54\,m                & 3570\,m           & 0 m\\
Cutoff            & 2400\,MV            & 810\,MV              & 2360\,MV             & 4500\,MV          & 6270\,MV \\
\tableline
\end{tabular}
\caption{Main characteristics of the NM stations used in this work \cite{Vaisanen2021}. \label{Tab::NMStations}}
\end{center}
\end{table*}
NM detectors, in spite of an excellent time resolutions, are unable to measure energy or particle type of the primary GCRs.
Thus the NM data have been combined with an appropriated model for their response and for the GCR flux variation. 
An important input for these calculations is the GCR proton spectrum in interstellar space (IS), $J^{\rm IS}_{p}$.
Here we have calculated the IS proton flux using an improved model of GCR acceleration and transport
based on our recent works\,\cite{Tomassetti2018PHeVSTime,Tomassetti2019Numerical,Tomassetti2015TwoHalo,Tomassetti2015PHeAnomaly,Tomassetti2012Hardening,Feng2016}.
Using direct IS measurements from Voyager-1 and -2, in combination with high-energy data from
the AMS-02 experiment \cite{Cummings2016,Aguilar2015Proton,Aguilar2015Helium}, we derived tight constraints for the IS fluxes of GCR protons and light nuclei.

\section{Data Analysis and Results}  
\label{Sec::Analysis}                

To extract information on the GCR flux variability from NM rates, we have modeled their response.
The NM data consist in energy-integrated counting rates $\mathcal{R}_{\rm NM}(t)$
provided on monthly basis. The rates used here, after being corrected for atmospheric pressure, detection efficiency and dead times,
reflect the total fluxes of secondary particles generated by the interactions of
GCRs with the atmosphere \cite{Dorman2009,Ghelfi2016,Usoskin2011}.
For a NM detector $d$, the link between the counting rate $\mathcal{R}_{\rm NM}^{d}$
and top-of-atmosphere GCR fluxes $J_{j}$ (with $j=$\p,\,\He,\C,\Oxy) can be expressed by:
\begin{equation}\label{Eq::NMRate}
  \mathcal{R}_{\rm NM}^{d}(t) = \int_{0}^{\infty}  dE \cdot \sum_{j={\rm GCRs}} \mathcal{H}^{d}_{j}(E)\cdot\mathcal{Y}^{d}_{j}(t,E)\cdot J_{j}(t,E)
\end{equation}
where $\mathcal{H}^{d}$ is the transmission function of GCR in the Earth's magnetic field, 
described as a smoothed Heavyside function across the geomagnetic cutoff $\R^{d}_{C}$ \citep{SmartShea2005,Tomassetti2015XS}.
The function $\mathcal{Y}^{d}_{j}$ describes the energy and time dependence of the detector response for the $j$-type particle,
including atmospheric showering and instrumental effects. 
We describe the response function using a parametric form \citep{Tomassetti2017BCUnc}.
Finally, the factor  $J_{j}(t,E)$ represents the time-dependent fluxes of all GCR species in proximity of Earth.
Here we consider the main species $j=$\,\p, \He, \C, \Oxy,  which contribute for nearly $99\,\%$ of the GCR flux.
The IS spectra $J^{\rm{IS}}_{j}$ outside the heliosphere are highly time-independent, \ie, 
the flux variations of $J_{j}(t,E)$ inside the heliosphere are entirely due to the solar modulation effect.
Here we express the relation between $J$ and $J^{\rm{IS}}$ using the Force-Field Approximation (FFA) model,
which is suitable for GCRs at the 10\,GeV energy scale, \ie, the typical scale of the NM data. 
Within the FFA, the modulated GCR flux at epoch $t$ is related to its LIS by the modulation potential $\phi$: 
\begin{equation}\label{Eq::ForceField} 
J(t,E) = \frac{(E+ m_{p})^{2}- m_{p}^{2}}{\left( E+m_{p} +\frac{|Z|}{A}\phi \right)^{2}-m_{p}^{2}} \times J^{\rm IS}(E + \frac{|Z|}{A}\phi(t))
\end{equation}
where $Z$ and $A$ are the charge and the mass number of the GCR particle, and $m_{p}$ is the proton mass.
The FFA can be interpreted as a shift in kinetic energy with the average value $\Delta{E}=E^{\rm IS}-E$, for
GCR particles, where $\Delta{E} - \frac{|Z|}{A} \phi$.
Within the FFA model, we convert the monthly average NM rates $\mathcal{R}_{\rm NM}(t)$ into time-series of modulation potential $\phi=\phi(t)$.
From this procedure we can compare data from different NM stations and extract information on the GCR flux variation.
To extract the information, we modeled the response of NM detectors as in \citep{Tomassetti2017BCUnc}. 
For each NM station $d$ at epoch $t$, the corresponding parameter $\phi^{d}(t)$ is obtained from
a fit of \emph{measured rate} $\hat{\mathcal{R}}^{d}$ with the \emph{calculated rate} $\mathcal{{R}}^{d}$ using Eq.\,\ref{Eq::NMRate},
together with the requirement that $\int_{\Delta T^{d}}{\mathcal{R}}^{d}(t) dt = \int_{\Delta T^{d}} \hat{\mathcal{R}}^{d}(t) dt$
over the total observation periods $\Delta T^{d}$ \citep{Tomassetti2017BCUnc}.
 
\begin{figure*}[!t]
\centering
\includegraphics[width=0.92\textwidth]{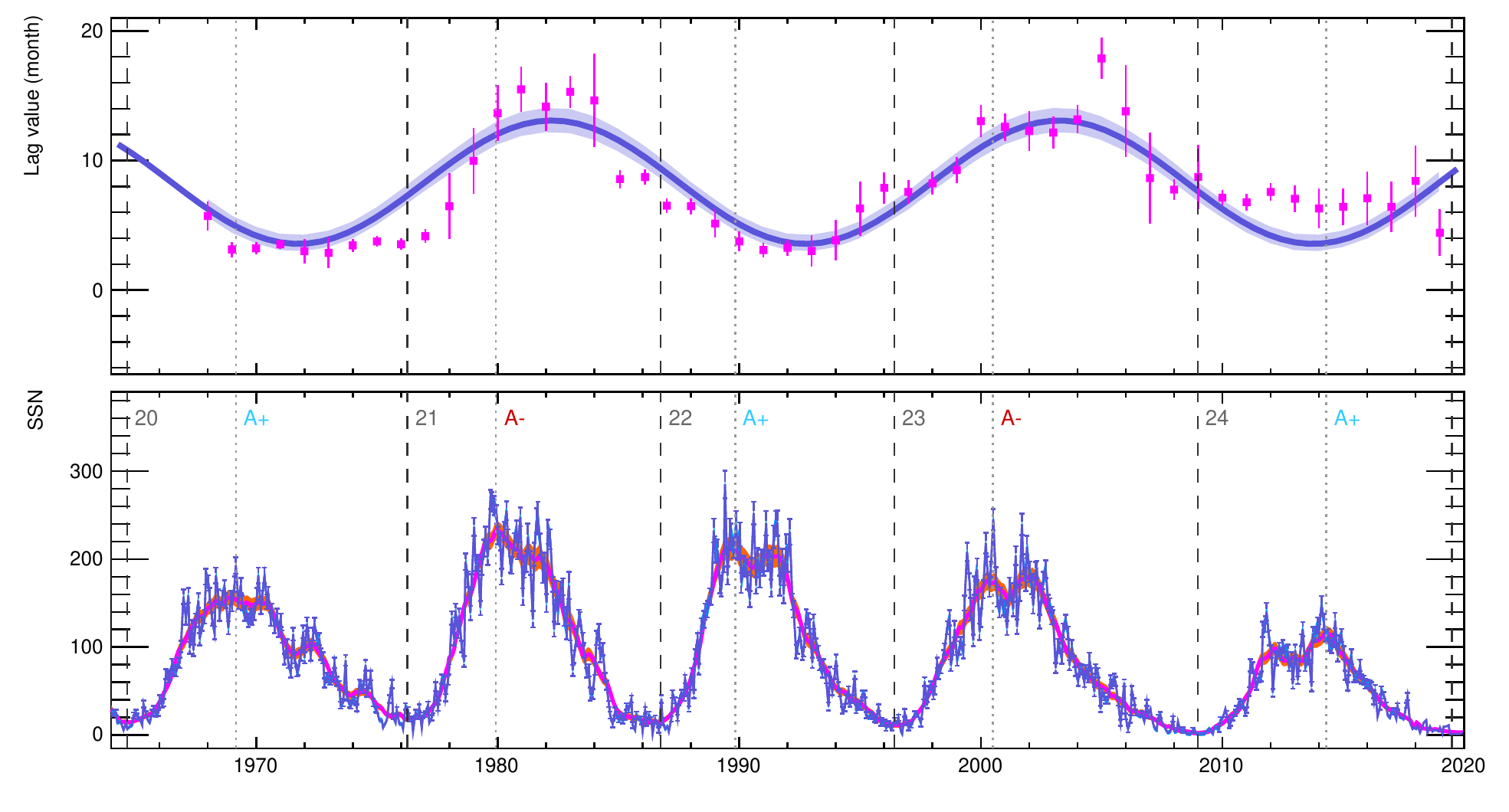}
\caption{\footnotesize{%
    Top: Temporal evolution of the modulation lag between 1965 and 2020 for the NM station in Rome, Italy.
    The continuous line is a sinusoidal fit giving period $T_{0}=21.6$\,months and average value $\tau_{0}=7.8$\,months.
    Bottom: evolution of the monthly and smoothed SSN, along with solar minima (dashed) and maxima (dotted) which
    define the solar cycle numbers (20 to 24) and their magnetic polarities $A+/A-$.
}}
\label{Fig::ccTimeLagVSTime2021}
\end{figure*}

The lag \DT{} between GCR and solar data is determined by means of a correlative analysis
between the NM-driven modulation parameter $\phi(t)$ and the smoothed SSN time-series $\hat{S}(t-\DT)$ in a given time range.
For a given station $d$, the modulation lag \DT{} is determined as the parameter that maximizes 
the correlation between the modulation  $\phi(t)$ at epoch $t$ and the smoothed SSN $\hat{S}(t-\DT)$.
A similar approach was adopted in other works \cite{Ross2019,Zhu2018,Iskra2019}.
To quantify the correlation level, we used the Spearman's rank correlation criterion.
To determine the temporal evolution of the lag, we have divided the total observation time into several time-windows.
We then accounted for all relevant sources of uncertainties such as those related with the experimental measurements,
the theoretical uncertainties in the IS flux models, the uncertainties on the smoothed SSN variance, and those
associated with the correlation coefficient.
The results are shown Fig.\,\ref{Fig::ccTimeLagVSTime2021}. The top panel, the reconstruction of the modulation lag
 between 1964 and 2019 (\ie, five solar cycles) using the monthly rates of the Rome NM station is shown.
From the figure, it can be noticed that the reconstructed time-series show a remarkable periodical behavior.
The data are fit with a sinusoidal function (solid red line) oscillating around an average value $\tau_{M}$:
\begin{equation}\label{Eq::TimeLagVSTime}
\tau(t) = \tau_{M} + \tau_{A} \cdot \cos\left[ \frac{2\pi}{T_{0}}\left( t - t_{P}\right)\right] \,,
\end{equation}
where the parameter $\tau_{A}$ represents the maximum amplitude of its variation,
$T_{0}$ is the oscillation period, and the reference time $t_{P}$ sets the phase.
From the time-series in the figure, one obtains an average lag $\tau^{0}=7.7{\pm}0.5$\,months and a period $T_{0}=21.6{\pm}0.8$\,years.
The average lag is consistent with other analyses based on large periods \citep{Tomassetti2017TimeLag}.
However, these results show that the lag is subjected to a quasiperiodical behavior, with a periodicity $T_{0}$ consistent with the 22-year
cycle of the Sun's magnetic polarity.
It is also interesting the comparison with the SSN evolution of the bottom panel, which shows the solar cycles 20 to 24
and the magnetic polarities $A^{+}/A^{-}$, where the polarity $A$ is defined as the sign of the Sun's $B$-field
outgoing from its North pole. At every solar maximum, the polarity is subjected to reversal, for a total periodicity of 22 years.

\section{Acknowledgements}  
\label{Sec::Results}        

We acknowledge the support of the Italian Space Agency ASI under agreement \emph{ASI-UniPG 2019-2-HH.0}.


\end{document}